\input ./style/arxiv-general.cfg
\documentclass[aoas,MSNbibl,nameyear,seceqn,dvips]{arximspdf}
\makeatletter
   \@ifpackageloaded{graphicx}{}{\usepackage{graphicx}}
\makeatother

\doi{10.1214/15-AOAS869}
\volume{9}
\issue{4}
\pubyear{2015}
\firstpage{2090}
\lastpage{2109}
\docsubty{FLA}

\begin{document}
\begin{frontmatter}

\title{``Virus hunting'' using radial distance weighted~discrimination\thanksref{T1}}
\runtitle{Radial DWD}

\begin{aug}
\author[A]{\fnms{Jie}~\snm{Xiong}\corref{}\ead[label=e1]{xiongj@unc.edu}},
\author[B]{\fnms{D. P.}~\snm{Dittmer}\ead[label=e2]{dirk\_dittmer@med.unc.edu}}
\and
\author[C]{\fnms{J. S.}~\snm{Marron}\ead[label=e3]{marron@unc.edu}}
\runauthor{J. Xiong, D. P. Dittmer and  J. S. Marron}
\affiliation{University of North Carolina at Chapel Hill}
\address[A]{J. Xiong\\
Department of Statistics\\
\quad and Operations Research\\
University of North Carolina\\
\quad at Chapel Hill\\
North Carolina 27599-3260\\
USA\\
\printead{e1}}
\address[B]{D. P. Dittmer\\
Lineberger Comprehensive Cancer Center\\
University of North Carolina\\
\quad at Chapel Hill\\
450 West Drive, CB\# 7295\\
Chapel Hill, North Carolina  27599-7295\\
USA\\
\printead{e2}}
\address[C]{J. S. Marron\\
Department of Statistics\\
\quad and Operations Research\\
University of North Carolina\\
\quad at Chapel Hill\\
North Carolina  27599-3260\\
 USA\\
\printead{e3}}
\end{aug}
\thankstext{T1}{Supported by public health service Grants
CA019014, and AI107810 to DPD.}

%
\received{\smonth{5} \syear{2014}}
%
\revised{\smonth{8} \syear{2015}}

\begin{abstract}
Motivated by the challenge of using DNA-seq data to identify viruses in
human blood samples, we propose a novel classification algorithm called
``Radial Distance Weighted Discrimination'' (or Radial DWD). This
classifier is designed for binary classification, assuming one class is
surrounded by the other class in very diverse radial directions, which
is seen to be typical for our virus detection data. This separation of
the 2 classes in multiple radial directions naturally motivates the
development of Radial DWD. While classical machine learning methods
such as the Support Vector Machine and linear Distance Weighted
Discrimination can sometimes give reasonable answers for a given data
set, their generalizability is severely compromised because of the
linear separating boundary. Radial DWD addresses this challenge by
using a more appropriate (in this particular case) spherical separating
boundary. Simulations show that for appropriate radial contexts, this
gives much better generalizability than linear methods, and also much
better than conventional kernel based (nonlinear) Support Vector
Machines, because the latter methods essentially use much of the
information in the data for determining the shape of the separating
boundary. The effectiveness of Radial DWD is demonstrated for real
virus detection.
\end{abstract}

\end{frontmatter}

\section{Introduction}\label{sec1}

A current major scientific challenge is the detection of viruses in
human blood samples. Cogent examples include HIV, the cause of AIDS;
poliovirus, which was considered eradicated, but has now emerged in
Syria and the Middle East; or middle east respiratory syndrome (MERS),
which entered the United States in May 2014 via a 44-year-old male who
traveled from Jeddah, South Africa, to Orlando, Florida, via London. At
home he developed fever, chills and a slight cough. He was admitted to
the hospital and later diagnosed with the MERS coronavirus. Since May
9, 2014, the World Health Organization (WHO) reported 536
laboratory-confirmed cases of MERS, including 145 deaths [\citet{WHO2014}].

For an effective treatment, a rapid and accurate detection of the
source of viral infection is crucial. The recent advent of deep DNA
sequencing techniques has led to a potentially powerful approach and it
gives rise to a new type of classification (discrimination) challenge.

A useful data space for virus detection comes from the DNA sequence and
alignment process, where virus-positive (the $+1$ class) and
virus-negative (the $-1$ class) samples are sequenced and the sequenced
DNA reads from each sample are aligned to a target virus. Reviews of
the DNA-sequence techniques can be found in \citet{Goldstein2013},
\citet{Mwenifumbo}, \citet{Rehm} and \citet{Grada}. A data vector counting the number of reads aligned to each
nucleotide position on the target virus is obtained for each sample.
Thereafter, data vectors from the 2 classes form the training set.


Because DNA sequencing is done at the base pair level of resolution,
the read depth vectors are quite long. As the number of samples is
relatively smaller, this analysis lies in the domain of high-dimension
low-sample size (HDLSS) data, which is an active research area where
the dimension $d$ of the data vectors is larger than the sample size
$n$; see, for example, \citet{Hall2005}, \citet{Liu2008}, \citet{JungandMarron2009}, \citet{FanandLv2010}, \citet{ShenShenandMarron} and \citet{YataandAoshima2013}. In this paper we focus on
HDLSS binary classification problems; see \citet{Marron2007}
and \citet{JiangMarronetal2009} for some examples.

An important aspect of DNA sequencing is that increasing the total
number of reads generated from a sample inflates entries of a data
vector. In many cases, the number of reads generated from samples
differs either due to the sequencing platform or to experimental
settings; see, for example, in \citet{Metzker}. The fact that we usually
collect different numbers of reads for the samples is regarded as a
\emph{bench effect} here. Bench effects may negatively impact the
classification and need to be handled properly. This is done here by
normalizing each data vector by dividing the entries by the \emph{L1
norm}, which is simply the summation of the entries in that vector,
since they are nonnegative. Therefore, normalized vectors all have
unit L1 norms to control for the bench effect.

Instead of normalizing by the L1 norm, other methods could be used to
adjust for the bench effect, for example, normalizations based on the
companion human genome. Such methods would be attractive in situations
where the goal was to determine the amount of virus present. However,
here our goal is to determine the presence or absence of virus (and, in
particular, we are focusing on trying to find rather small amounts). In
this context, our normalization seems the most powerful.

A consequence of this L1 normalization is that the normalized data
vectors can be geometrically represented as points on the standard unit
simplex. Data points with more nonzero entries lie more toward the
interior of the unit simplex. When all entries are approximately the
same, the data point is near the center. On the contrary, the more
zeros in a vector, the closer this data point is to one of the vertices
of the unit simplex. In the extreme case with only one nonzero entry
``1'' in the vector, the data point is at a vertex.

\begin{figure}[b]

\includegraphics{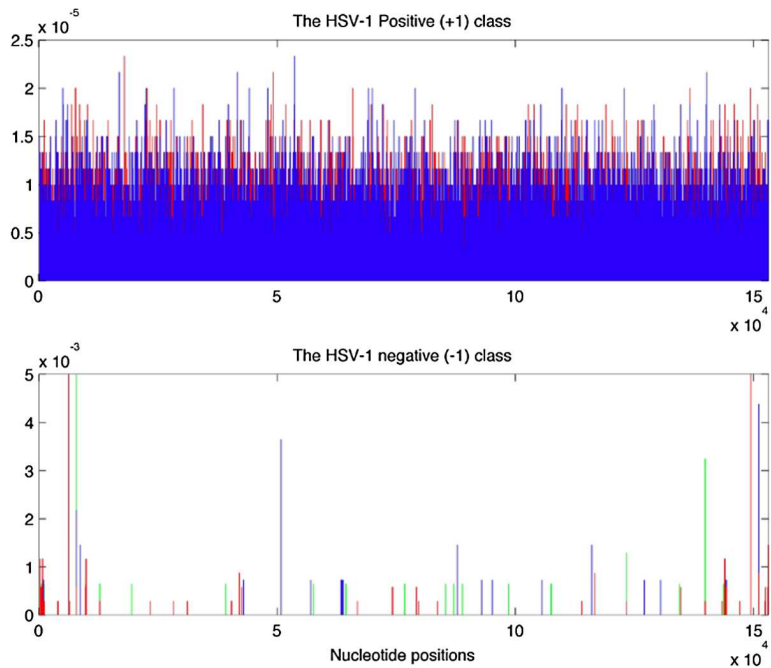}
\caption{Overlaid plot of 2 normalized data vectors from the HSV-1
positive (the $+1$) class in the top panel and 3 data vectors from the
HSV-1 negative (the $-1$) class in the lower panel, all with different
colors. The overall entries of the $+1$ data vectors are relatively small
and have quite comparable amplitudes, while the entries of the $-1$ data
vectors have ``spikes'' (which are located at quite divergent positions).}\label{fig1}
\end{figure}

Figure~\ref{fig1} shows how different the virus positive and virus negative
samples are, and motivates exploiting simplex geometry, by showing an
overlaid plot of normalized data vectors from an HSV-1 (a human
herpesvirus) detection problem. HSV-1, or human herpesvirus-1, is the
leading cause of nontraumatic blindness and can cause fatal
encephalitic disease in children. The virus can be treated with
acyclovir, if and only if diagnosed rapidly and accurately. Both serum
and cerebral spinal fluid are used for diagnosis and can be readily
obtained for sequencing. In Figure~\ref{fig1}, we overlaid 2 (out of 8) data
vectors for the HSV-1 positive (the $+1$) class (top panel) and 3 (out of
24) data vectors from the HSV-1 negative (the $-1$) class (lower panel).
The overall entries of the positive data vectors are relatively small
and have relatively comparable amplitudes (top panel). The nonzero
entries of the negative data vectors are very sparse and have much
larger amplitudes (about 200 times larger than that of the positive
samples, lower panel of Figure~\ref{fig1}). This is a property of all virus
detection problems, since the negative sequences are chosen to be
genetically very different from the virus. The aligned reads (from the
negatives to the virus sequence), on the other hand, are often short
stretches of sequence which are of reduced complexity, that is, repeats
or single nucleotide (either A, C, T or G) runs.

\begin{figure}[b]

\includegraphics{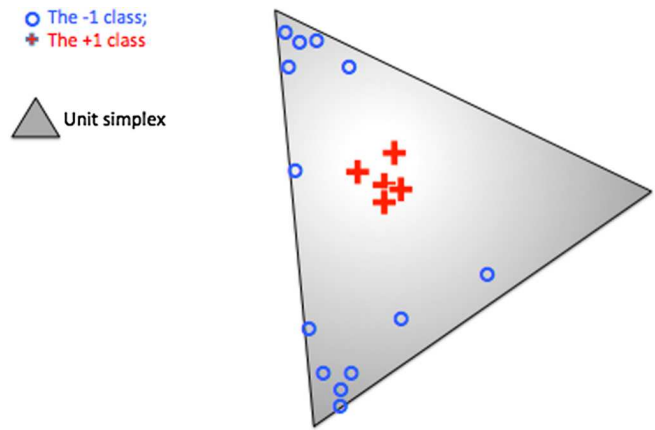}
\caption{A simplified example of normalized data vectors of the
positive ($+1$ class) data (red plus signs) and the negative ($-1$ class)
data points (blue circles). The unit simplex is shown as the gray
triangle. Because there are many zeros in the $-1$ data vectors, they
typically locate at the vertices of the unit simplex while the
positives are closer to the center.}\label{fig2}
\end{figure}

If we keep the same range of $y$-axis in plotting the positive data
vectors as in plotting the negative data vectors, one can see almost
nothing since the amplitudes of the former ones are much smaller than
the latter ones. Equivalently speaking, the positives are close to the
center of the simplex, while the negatives lie near to a diverse set of
vertices of the unit simplex, because the differently colored spikes of
the negative data vectors are located at quite divergent positions. A
simple model for data on the unit simplex is given in Figure~\ref{fig2}, where
the 3-$d$ unit simplex is shown as a gray triangle while some $+1$ ($-1$)
class data are shown as red plus signs (or blue circles, resp.).
It is not hard to see that linear methods will struggle to capture the
differences between classes in this case.

In our work, the DNA alignment data vectors are often of dimension
100{,}000 to 200{,}000, while the sample size is usually much smaller.
Equivalently, data can be seen as points on the high-dimensional unit simplex.

Figures~\ref{fig1} and \ref{fig2} suggest that the $-1$ class departs from the center
of the simplex (where the $+1$ class lies) in many diverse directions so
that the theoretical Bayes classification boundary (assuming a
probability distribution for each class) is highly nonlinear. Note that
the discrimination in radial directions appears to be attractive. This
motivates the development of Radial DWD in order to incorporate such a
nonlinear pattern. As detailed in Section~\ref{sec4}, by optimizing a
hypersphere over its center and radius, Radial DWD separates the 2
classes, favoring putting the $+1$ ($-1$) class inside (outside) the
hypersphere. The computation of Radial DWD through solving a sequence
of Second Order Cone Programs \cite{AlizadehandGoldfarb2003} is carried
out by an interior point optimization package called SDPT3, developed
by \citet{TTT2001}. A future sample will be classified as
$+1$ ($-1$) when it is located inside (outside) the hypersphere.

\begin{figure}

\includegraphics{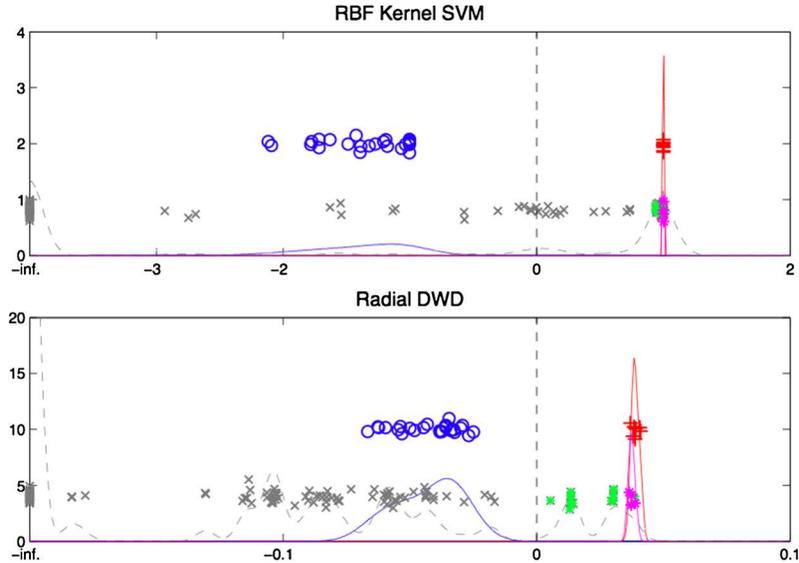}

\caption{An HSV-1 classification example to compare the performance of
RBF kernel SVM and Radial DWD. Trained on red plus signs versus blue
circles, the former method endures a high false positive error since
many negative test samples (gray $x$-symbols) are on the same side of the
separating boundary as the positive class, while Radial DWD
successfully classified all positive HSV-1 samples (magenta asterisks)
and related viruses (green asterisks) with no false positive.}\label{fig3}
\end{figure}

A standard approach to HDLSS classification problems is linear methods,
such as Mean Difference [MD, \citet{ScholkopfandSmola2002}],
penalized logistic regression with LASSO penalty [LASSO, \citet{Tibshirani1996}], Support Vector Machine
[SVM, \citet{Vapnik1995}, \citet{Shawe-TaylorCristianini2004}] and Distance Weighted Discrimination
[DWD,  \citet{Marron2007}]. Figure~\ref{fig2} suggests that, as the dimension and diversity
of the $-1$ class grow, such methods will be severely inefficient. This
issue is carefully studied for actual ``virus hunting'' in
Section~\ref{sec2}
and by simulation in Section~\ref{sec3}. It is natural to wonder if a more
serious competitor to Radial DWD is a nonlinear kernel Support Vector
Machine classification [\citet{Burges1988},
\citet{HTF2009}]. The most popular of these is the Radial Basis Function (RBF)
kernel. The virus detection capability of these methods are compared in
Figure~\ref{fig3}, where RBF kernel SVM and Radial DWD classification are
illustrated. Note that in the machine learning literature, RBF is a
synonym for ``Gaussian kernel.''

In Figure~\ref{fig3}, RBF kernel SVM and Radial DWD are trained using 8 HSV-1
positive (red plus signs) and 24 HSV-1 negative (blue circles) data
vectors, which are partially shown as an overlaid plot in Figure~\ref{fig1}.
Signed distances to the corresponding separating boundary (the black
vertical dashed line) are depicted along the $x$-axis as a jitter plot.
Random heights are assigned in order to visually separate the points.
Additionally, 127 new samples are used as a test set and kernel density
estimates are given for each group. While the majority of test samples
are shown as gray $x$-symbols, 4 are highlighted in magenta since they
are HSV-1 positive human samples; 14 are highlighted in green since
they are highly related herpesviruses (with nonhuman hosts). The
related viruses share significant sequence identity (traditionally
larger than 35\%) with the reference virus, but may infect animals
rather than humans. Domestic cats and cattle, for instance, can be
infected with a herpesvirus homologous to HSV-1.

The performance of RBF kernel SVM is far from satisfactory: although
positive samples (magenta and green asterisks) are very close to the
true positives, many (69) grays (unrelated samples) are also classified
as HSV-1 positive. This is expected since kernel methods require a type
of ``data richness,'' that is not present in the virus hunting problem.
In particular, they work well in situations where training data can be
found in all of the various regions where the test data will appear.
But in virus hunting data analysis, that completely breaks down.

Radial DWD shows a superior classification result not only because it
correctly classified all HSV-1 positive samples but also because the
positive samples are grouped reasonably well: HSV-1 positive human
samples (magenta asterisks) are tightly clustered with the positive
training data (red plus signs); related herpesviruses (green asterisks)
are clustered according to the host species that they infect---from
the right to the left---monkey, pig and cattle. The grouping property
of Radial DWD can be exploited to classify new viruses, for example, in
different animal hosts, as they would be related, but not identical to
the known ones.

For some data sets, it will be sensible to use a given point, for
example, the center of the simplex or the sample mean, as the
centerpoint of the separating sphere. Therefore, solving the associated
optimization problem will be generally easier. However, the center of
the simplex seems inappropriate for virus hunting, as due to various
biological effects, even in the limit as the number of reads goes to
infinity, the read depth vector is not flat. The sample mean can be
appropriate in many situations, but as the centroid classifier is often
a lot less efficient in many high-dimensional biological settings, we
expect Radial DWD to often be worth the overhead of the more complex
optimization. Furthermore, we also have our eye on generalizing to
other data types, where we believe the property of Radial DWD having
conventional DWD as a limit (as the center goes to infinity in a
particular direction, with the radius also growing) will become very important.

A full description of this HSV-1 classification is given in Section~\ref{sec2},
where we carefully compare Radial DWD with some linear and nonlinear
competitors and the superiority of Radial DWD under this radial context
is discussed in detail. A~similar conclusion can be drawn from the
simulation study in Section~\ref{sec3}. Radial DWD optimization and an iterative
algorithm to solve it can be found in Section~\ref{sec4}. An introduction to
virus detection, insights about the Dirichlet distribution and the
high-dimensional unit simplex, along with more details of our data sets
and some proofs, can be found in the supplementary materials in \citet{Xiong2015}.

\section{Virus detection data analysis}\label{sec2}

As briefly described in Section~\ref{sec1}, Radial DWD presents an appealing
virus detection capability. A broader comparison between Radial DWD and
its linear and nonlinear competitors is given in this section through
analyzing a real data example of detecting the $\alpha$-Human
Simplexvirus 1 ($\alpha$-HSV-1 or HSV-1). This virus is a subfamily of
Human Herpesvirus (HHV). The data set consists of the following 2 subsets:
\begin{itemize}
\item The training data are $n_{+}=8, n_{-}=24$ vectors of dimension
152{,}261, which is the DNA length of HSV-1. Entries of each data vector
correspond to the nucleotide positions in the virus DNA sequence. The
training data of the $+1$ (HSV-1 positive) and $-1$ (HSV-1 negative)
classes are normalized to the unit simplex (of dimension 152{,}261). The
$+1$ class tends to locate near the center while the $-1$ class tends to
locate near a diverse set of vertices of the simplex. Classifiers are
trained using the $+1$ versus the $-1$ classes.
\item The test set consists of the DNA alignment vectors from the
following samples: 4 HSV-1 positive human samples (not appearing in the
training), 14 nonhuman $\alpha$ Simplexvirus-1 (including 5 monkey
Simplexvirus-1, 8 pig Simplexvirus-1 and 1 cow Simplexvirus-1) and 109
much more distantly related viruses.
\end{itemize}

Nonzero data vectors are normalized to the unit simplex and can be
viewed as points on it. Samples whose DNA alignment vectors are zero
vectors are put at $-\infty$. This is reasonable since zero vectors
only exist in the $-1$ class training set or the test set: (a) if the
sample comes from the $-1$ class training set, it has no effect on the
calculation of the separating sphere (interpreting the reciprocal of
$-\infty$ to be zero); (b) if the sample comes from the test set, it
should surely be classified as $-1$ and $-\infty$ is viewed as outside
the separating hypersphere.

\begin{figure}[b]

\includegraphics{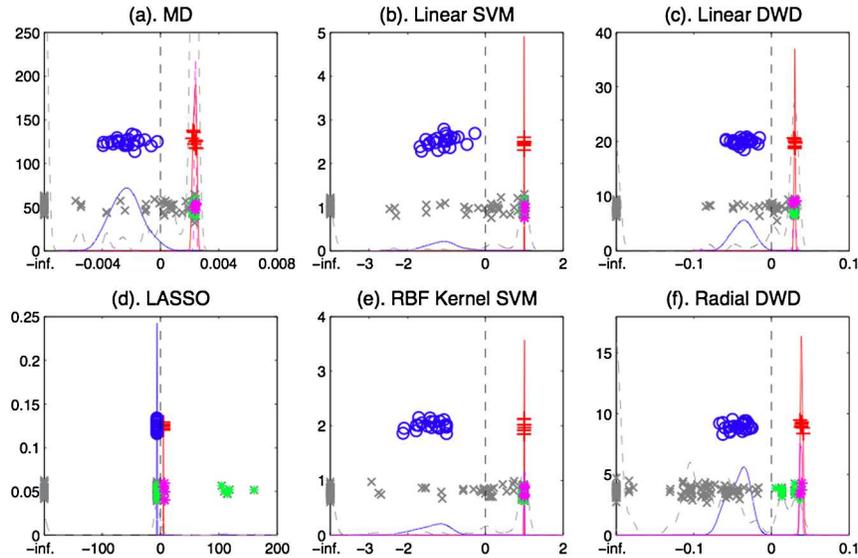}

\caption{Real data example of an HSV-1 classification problem. We show 6
panels of 1-dimensional ``signed distance to separating boundary''
plots to compare Radial DWD [panel (\textup{f})] with MD, LASSO, linear DWD, SVM
and RBF SVM. Red plus signs are $+1$, blue circles are $-1$, magenta
asterisks are HSV-1 positive humans, green asterisks are related
nonhuman herpesviruses, gray $x$-symbols are nonpositive samples. Figure~\protect\ref{fig4} shows the superior performance of Radial DWD.}
\label{fig4}
\end{figure}

The classification performance of Radial DWD is compared with a number
of popular classification methods including MD, LASSO, linear DWD, SVM
and RBF Kernel SVM in Figure~\ref{fig4}. Quadratic SVM gave results that were
quite similar to RBF SVM, so it is not shown here. For methods
including a separating plane, relative performance comes from the
projection onto the normal vector, shown as the horizontal axes in
Figure~\ref{fig4}. Radial DWD is similarly interpreted as the signed distance to
the separating sphere. The $+1$ training data are shown as red plus
signs, $-1$ training data as blue circles, HSV-1 positive human samples
(real human DNA samples that are infected by HSV-1) as magenta
asterisks, related $\alpha$ simplex herpesviruses as green asterisks
and other samples as gray $x$-symbols (known to be HSV-1 negative). The
position of the separation boundary is shown by the black vertical
dashed line. Signed distances to the separating boundaries are depicted
along the horizontal axis, while the vertical perturbation is used for
visual separation of the points. Kernel density plots are provided as well.

Panels (e) and (f) are the same as in Figure~\ref{fig3} (except that the
length-width ratio of the figures are different) and the superior
performance of Radial DWD is explained there. While the training data
is well separated in all cases in Figure~\ref{fig4}, the good classification
property may not carry over to the test samples. The performance of MD,
SVM, DWD and RBF SVM tend to be similar in this example where the false
positive rates are very high (larger than 50$\%$), that is, most of the
negative gray $x$-symbols are to the left of the dashed line. This
performance contrasts sharply with panel (f) where all gray $x$-symbols
are to the right. Meanwhile, LASSO presents a unique behavior with zero
false positive. However, it fails to correctly classify 8 (out of 14)
HSV-1 related viruses (green asterisks) since they fall on the
left-hand side of the LASSO-separating hyperplane. The other 6 HSV-1
related viruses are much further from the positive training data (red
plus signs) to the right. Our simulations show that LASSO tends to pick
out a small subset of nucleotide positions and classify data merely
based on very limited information gained on those positions, which
results in poor classification.

An additional insightful comparison of methods using simulated data
sets appears in Section~\ref{sec3} and Radial DWD will be shown to have a much
better classification accuracy in terms of both lower false positive
and lower false negative error rates, while all the other competitors
considered here perform poorly. Note that real data examples of $\beta$-HHVs and $\gamma$-HHVs (the other 2 subfamilies of HHV)
classification were also analyzed and examples can be found in the
supplementary materials in \citet{Xiong2015}.

In addition to giving outstanding classification results when one class
is widely distributed around the other, the computation of Radial DWD
is fast enough to be useful for modern scale bioinformatics data sets.
The computational speed is nearly independent of the dimension of the
data vectors because the method is based on a QR decomposition (see
Section~\ref{sec4} for detail). In particular, the full set of simulations shown
in Section~\ref{sec3}, involving many replications, was done in a few hours.

\section{Simulation study}\label{sec3}

Section~\ref{sec2} showed that Radial DWD outperforms its linear and nonlinear
competing classifiers for real virus detection data, and this idea is
further emphasized in this section by a simulation study. Our
simulations are based on Dirichlet distributions which are a popular
and broad family of distributions on the unit simplex. Figure~\ref{fig5} shows
the classification results, aimed at modeling the behavior observed in
real data in Figure~\ref{fig4}, detailed in Section~\ref{sec31}. Broader simulation
results are discussed in Section~\ref{sec32}.

\subsection{Simulation 1}\label{sec31}

\begin{figure}[t]

\includegraphics{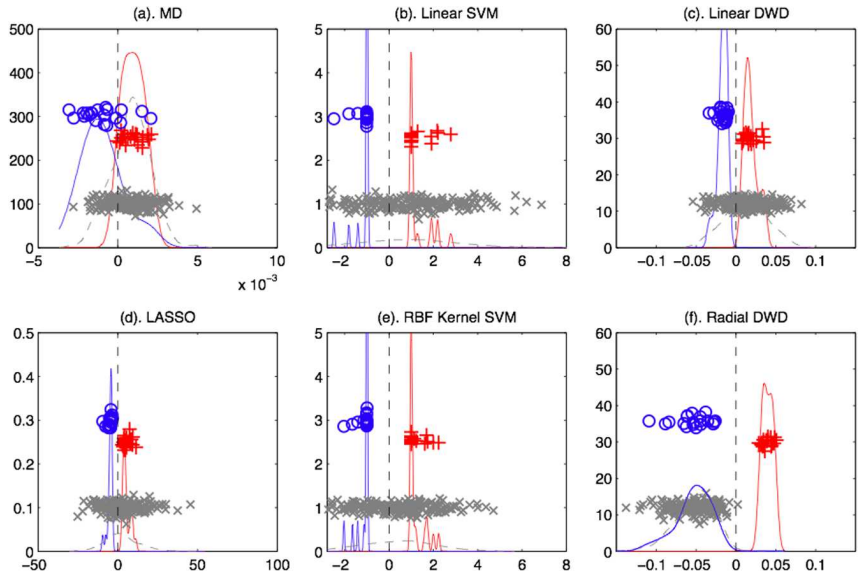}

\caption{A simulated example illustrating the potential for improved
performance of Radial DWD, in the spirit of Figure~\protect\ref{fig4}. Class $+1$ is shown
as red pluses, $-1$ as blue circles, test samples as gray $x$ symbols. The
vertical axis shows random heights to visually separate the points,
along with kernel density estimates (i.e., smooth histograms). The
separating boundaries are calculated using the following: \textup{(a)} MD, \textup{(b)}
Linear SVM, \textup{(c)} Linear DWD, \textup{(d)} LASSO, \textup{(e)} RBF SVM and \textup{(f)} Radial DWD.
Except Radial DWD, all the other methods have poor classification
performance for the test samples, which should be mostly to the left of
the dashed line in each case.}\label{fig5}
\end{figure}

The simulated data in Figure~\ref{fig5} have dimension $d=50$, with $n_{+}=20$
class $+1$ samples represented as red plus signs and $n_{-}=20$ class~$-1$
samples represented as blue circles. Data are simulated using the
Dirichlet distribution $\operatorname{Dirichlet}(\alpha)$, supported on the
unit simplex.

The parameter $\alpha\in R^{d+}$ determines the mode and dispersion of
the Dirichlet distribution. If all the entries in $\alpha$ are the
same, the distribution is centered on the unit simplex. Suppose the
common entries are larger (less) than 1, increasing (decreasing) the
entries makes the distribution more concentrated to the center
(vertices) of the unit simplex; suppose the common entries are exactly
1s, the corresponding distribution is the uniform on the simplex.
Examples in 3 dimensions can be found in the supplementary materials in
\citet{Xiong2015}.

The $+1$ class data in Figure~\ref{fig5} are drawn from $\operatorname{Dirichlet}(\alpha
_{+})$ with $\alpha_{+} =(5,\ldots,5)$ and the $-1$ class data are
generated from $\operatorname{Dirichlet}(\alpha_{-})$ with $\alpha_{-} =(0.5,\ldots,0.5)$.

Classifiers, including MD, LASSO, linear SVM, RBF Kernel SVM, DWD and
Radial DWD, are trained on the red pluses and blue circles. We assess
the performance by classifying 200 new test samples drawn from the $-1$
class population. The test samples are shown in Figure~\ref{fig5} as gray
$x$
symbols. Note that the Quadratic Kernel SVM (QSVM) was also considered.
It performed very similarly with RBF SVM in this particular example,
and hence is not shown here.

In each panel of Figure~\ref{fig5}, the signed distances of the data points to
the optimal separating hyperplane are shown on the horizontal axes. The
position of each separating hyperplane is shown as a dashed line. Data
points that fall on the same side of the hyperplane as the $+1$ ($-1$)
class will have positive (negative) distances. Kernel density plots
(e.g., smooth histograms) are provided as another way of viewing the
population of each class. As shown in Figure~\ref{fig5}(a), MD performs poorly
(with many gray test points to the right of the boundary) since the
separation of classes in this example is not a shift of means. In
particular, 152 out of 200 samples are misclassified as $+1$. Figure~\ref{fig5}(b)
shows that the 2 training classes are linearly separable by using SVM,
but the training data from both classes pile up at the margin.
Moreover, 129 out of 200 test samples are misclassified as $+1$.

Data piling is a sign of overfitting and is very undesirable since the
corresponding separating hyperplane is driven heavily by the particular
realization of the data at hand [see \citet{Marron2007}]. DWD
was developed to address this ubiquitous problem with SVM, yet Figure~\ref{fig5}(c) is similar to (b). The phenomenon of data piling is diminished as
expected from the ideas of \citet{Marron2007}. However, the
performance of DWD for this test set is far from satisfactory because
radial separation is the key: again, many (142 out of 200) test samples
are misclassified as $+1$.

Classification using LASSO is illustrated in Figure~\ref{fig5}(d). The training
data are well separated, but 117 of the 200 test samples are
misclassified. Figure~\ref{fig5}(e) shows the classification using the RBF
(nonlinear) Kernel SVM. When the training set is linearly separable,
kernel SVM behaves like the linear counterpart but may overfit the
training data more severely under HDLSS assumptions. Although the
dimension is fairly moderate, data piling still exists. The expected
improvement over the linear counterpart is present in the sense that
only 120 out of 200 test samples are misclassified, although this is
still unacceptably poor.

A much improved performance and classification accuracy can be observed
in Figure~\ref{fig5}(f) where Radial DWD is applied. Training data are well
separated with no signs of data piling and, except for one test data
point, all the other test samples are correctly classified, showing
that Radial DWD solves the overfitting problem one may intuitively
expect from the RBF kernel SVM in HDLSS radial contexts.

It is not surprising that, despite the underlying nonlinear pattern,
SVM and DWD successfully separate the 2 training classes due to the
large size of the data space. However, the good classification
performance does not carry over to the test samples, which may differ
from the $+1$ class in directions that do not appear in the $-1$ class
training data. This highlights the limitation of linear methods in this
type of context. Figures~\ref{fig4} and~\ref{fig5} together make it clear that the
intuitive ideas in Section~\ref{sec1} are indeed the drivers of the observed
superior performance of Radial DWD. Thus, simulating data from the
Dirichlet distribution is useful and insightful to understand the data
structure of the virus discovery.

\subsection{Simulation 2}\label{sec32}

Next a broader simulation study is conducted. The training data and the
test data are simulated on the unit simplex using $\operatorname{Dirichlet}(\alpha)$
with $\alpha$ summarized in Table~\ref{parset}. In each example, $n_{+}=20$ $+1$
class and $n_{-}=50$ $-1$ class data of dimension $d = 10, 50, 100, 500,
1000, 5000,\break 10{,}000, 50{,}000, 100{,}000$ are generated in order to cover a
range from non-HDLSS to extreme HDLSS cases. Additionally, in panels
(a1) and (a2), 5000 test samples are drawn from the $-1$ class in order
to assess the false positive rate; in panels (b1) and (b2), 5000 test
samples are drawn from the $+1$ class in order to assess the false
negative rate. Thirty repetitions are done for each case and each dimension.

\begin{figure}[t]

\includegraphics{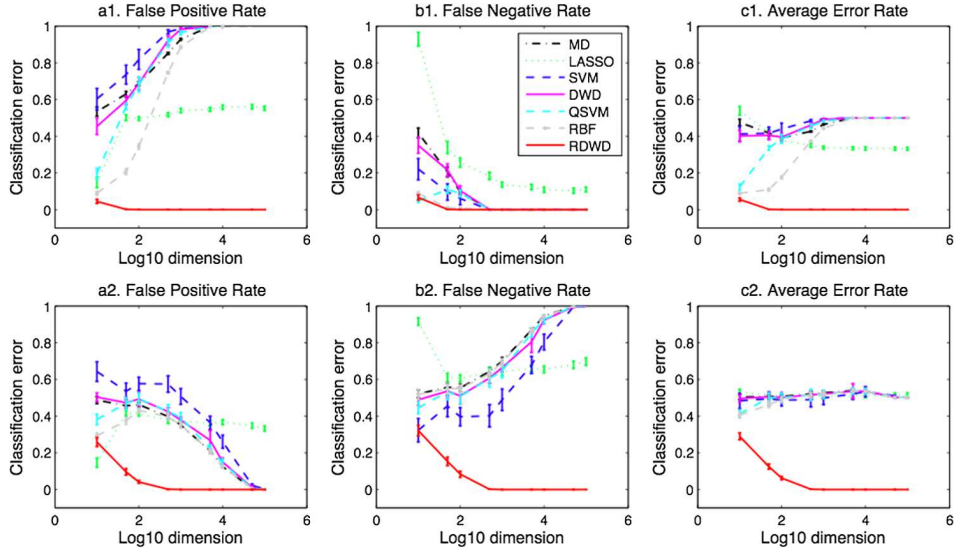}

\caption{A simulation study illustrating the potential for improved
performance of Radial DWD. The false positive rate is depicted in
panels \textup{(a1)} and \textup{(a2)} under each parameter setting, with the
corresponding false negative rate (under the same training setup) in
\textup{(b1)} and \textup{(b2)}. The average of the false positive and false negative
rate is shown in panels \textup{(c1)} and \textup{(c2)}, respectively. Classification
error is calculated for the following: $-\cdot$ MD $\cdot\cdot$ LASSO,
$--$ Linear/Quadratic/RBF SVM, $-$ Linear DWD and $-$ Radial DWD (RDWD).
A color key is also given. Error bars are obtained by repeating the
simulation 30 times for each dimension $d$. Figure~\protect\ref{fig6} shows the
outstanding performance of Radial DWD relative to typical methods in
these radial settings.}\label{fig6}
\end{figure}

The tuning parameters in LASSO, (linear/Quadratic/RBF) SVM, DWD are
determined by 5-fold cross-validation. Classifiers are trained using
the $+1$ versus the $-1$ class. Classification error (false positive and
false negative) is calculated for classifying the 5000 test samples and
is illustrated in Figure~\ref{fig6}.

%

In the first simulation in panels (a1), (b1) and (c1), the $+1$ class is
simulated uniformly on the  simplex using $\operatorname{Dirichlet}(1\cdots 1)$, while
the $-1$ class is simulated near the vertices of the simplex, as given in
Table~\ref{parset}. The class separation is hard in low dimensions, but, as
dimension grows, the relatively low sample size of the training data
makes the separation easier. It can be seen in panel (a1) that when
dimension is low (around 10), RBF kernel SVMs and Radial DWD perform
similarly well with false positive error rates below 10$\%$, LASSO and
Quadratic kernel SVM follows and all the other linear methods perform
poorly. As dimension goes to $\infty$, the false positive error of
Radial DWD shrinks to zero quickly, while that of the MD/SVM/DWD/RBF
kernel SVM/Quadratic kernel SVM goes to 1; that of LASSO converges to
around $50\%$. A quite different tendency can be observed in panel (b1)
when the false negative rate is being examined. When dimension is low,
LASSO tends to have a very large false negative error, but the error
shrinks to zero quickly as dimension grows, as do the false negative
error rates for the other methods. The average of the 2 types of errors
is summarized in (c1). It is not hard to see that the kernel SVMs and
Radial DWD are comparably good in low dimensions; the error rate of the
former one converges to around $50\%$, while that of the latter one
converges to zero quickly as dimension grows. Additionally, the average
error rate of MD/SVM/DWD is relatively stable (around $50\%$); the
average error rate of LASSO is around $35\%$ for large dimensions.

\begin{table}[b]
\tablewidth=300pt
\caption{Parameter $\alpha$ used in simulation}
\label{parset}
\begin{tabular*}{300pt}{@{\extracolsep{\fill}}lccc@{}}
\hline
\multicolumn{1}{@{}l}{\textbf{Case \#}} & $\bolds{+1}$ \textbf{class} & $\bolds{-1}$ \textbf{class} &
\textbf{Corresponding panels in Figure~\protect\ref{fig6}}\\
\hline
1 & $(1\cdots 1)$ & $(0.1\cdots 0.1)$ & (a.1) (b.1) and (c.1) \\
2 & $(1\cdots 1)$ & $(0.5\cdots 0.5)$ & (a.2) (b.2) and (c.2)\\
\hline
\end{tabular*}
\end{table}

The second simulation in panels (a2), (b2) and (c2) is similar to the
first except that the $-1$ class is closer to the center. This is even a
harder classification problem when dimension is low. An almost opposite
tendency could be observed in (a2) and (b2), compared to (a1) and (b1).
Except LASSO, the false positive rate [in (a2)] of all methods shrinks
to zero, while that of Radial DWD decreases much faster; the false
positive error rate of LASSO is around $35\%$ for large dimensions.
Shown in panel (b2), the false negative error rate of Radial DWD still
decreases to zero as dimension grows, however, the error rates of its
competitors goes to 1 (or above $60\%$ for LASSO). The average of the
false positive and the false negative rate is illustrated in panel (c2)
where a similar pattern as (c1) can be observed, except that even in
low dimensions, kernel SVMs did not work as well as Radial DWD. When
dimension is high, all Radial DWD's competing classifiers have error
rates around about $50\%$ (i.e., essentially random choice).

We also studied several other examples [see Supplement in \citet{Xiong2015}]. They show fairly similar results. As suggested by our current
simulations, Radial DWD outperforms MD, LASSO (linear, Quadra\-tic, RBF)
SVM and linear DWD when the radial separation is the key player to
discriminate classes. As noted before, the full set of simulations
shown in Section~\ref{sec3}, involving many replications, was done in a few hours.

Radial DWD performs well with this type of data because of the
particular geometry. In Figure~\ref{fig1}, we show that there are scaling issues
with these coverage vectors as data objects, which are handled by
dividing by the sum of the entries. This transformation means the data
live on the unit simplex, hence, we study its geometry. Furthermore,
because the dominant spikes in Figure~\ref{fig1} are in different locations, the
data negative samples are widely distributed around the simplex, in
many different directions. We tried to illustrate this phenomenon with
a grossly simplified (because human perception tends to fail beyond 3
dimensions) toy example in Figure~\ref{fig2}. But it is the major exaggeration
of this effect, that naturally occurs in this HDLSS context, that
drives the major breakthrough of Radial DWD relative to the existing
competitors (which were not designed for this setting).

\section{Radial DWD optimization}\label{sec4}
\subsection{Formulate the optimization problem}\label{sec41}

To set notation, let $n$ denote the number of training $d$-vectors
$x_i$ with corresponding class labels $y_i\in\{-1,+1\}, i=1\cdots n$. We
let $X$ denote the $d\times n$ matrix whose columns are $x_i$. Let $e$
denote an $n$-vector of 1s. Let $n_{+}=\sum^{n}_{i=1}I_{\{y_{i}=+1\}
}$ and $n_{-}=\sum^{n}_{i=1}I_{\{y_{i}=-1\}}=n-n_{+}$ be the sample
size of the $+1$ class and the $-1$ class, respectively. Denote $O\in
R^{d}$ as the center of a candidate separating sphere, and let $R\in
R^{+}$ be the radius, and define the signed residual of the $i$th
data point as $\bar{r}_i=y_{i}(R-\|x_{i}-O\|_{2})$, where $\|\cdot\|_2$
represents the Euclidean norm. We would like to search for $O$ and $R$
such that $\bar{r}_i$ are positive and large, which requires the $+1$
class to lie inside and the $-1$ class to lie outside the hypersphere.
However, in order to incorporate the case when the 2 classes are not
separable by a hypersphere, we allow classification error by adding
nonnegative ``slack variable'' $\varepsilon_{i}$, as in \citet{Burges1988}
and \citet{Marron2007}, and define perturbed residuals as
$r_i=y_{i}(R-\|x_{i}-O\|_{2})+\varepsilon_{i}$. We now define the
optimization problem for Radial DWD as follows:
%
\begin{eqnarray}
&& \operatorname{Min}_{r,O,R,\varepsilon} \ \sum_{i}
\frac
{1}{r_i}+Ce^{T}\varepsilon
\nonumber
\\[-2pt]
&&\mbox{s.t.}\qquad r_{i}=y_{i}\bigl(R-\|x_{i}-O
\|_{2}\bigr)+\varepsilon_{i},\qquad i=1,\ldots,n,
\\[-2pt]
&&\phantom{\mbox{s.t.}}\qquad R\ge0, \qquad r\ge0,\qquad \varepsilon\ge0,
\nonumber
\end{eqnarray}
where $r$ is a vector of $r_{i}$, $\varepsilon$ is a vector of
$\varepsilon_{i}$ and $r\ge0$ and $\varepsilon\ge0$ are in the
component-wise sense, and $C$ is the penalty parameter of
misclassification, as appears in SVM and DWD. It can be seen that the
influence of the $-1$ class data decreases as they get further away from
the separating hypersphere. The influence shrinks to zero for the $-1$
class data located at infinity. However, this is not true for the $+1$
class (because of the penalty term). Following \citet{Marron2007}, we
linearize the objective function by defining $\rho_{i}=(r_{i}+\frac
{1}{r_i})/2$ and $\sigma_{i}=(\frac{1}{r_i}-r_{i})/2$, so that $\frac
{1}{r_i}=\rho_{i}+\sigma_{i}$, $r_{i}=\rho_{i}-\sigma_{i}$.
Additionally, we relax the constraints
\begin{eqnarray*}
&& \bigl\{\rho^{2}_{i}-\sigma^{2}_{i}=1,
\rho_{i}-\sigma_{i}\ge 0, i=1,\ldots,n\bigr\}
\end{eqnarray*}
to the second order cone constraint
\begin{eqnarray*}
&& \bigl\{(\rho_{i},\sigma_{i},1)\in S_{3},
i=1,\ldots,n\bigr\},
\end{eqnarray*}
where the Second Order Cone of dimension $k$ is defined as
\begin{eqnarray*}
&& S_{k}=\bigl\{(\varsigma;\mu)\in R^{k}: \varsigma\ge\|
\mu\|_{2}\bigr\}.
\end{eqnarray*}
One can show that when the 2 classes are separable by using a
hypersphere, this relaxation will not change the optimal solution. By\vspace*{1pt}
the transformation of $\frac{1}{r_{i}}$ and the substitution with
Second Order Cone constraints, the optimization problem becomes
%
%
\begin{eqnarray}
&& \mathrm{Min}_{\rho,\sigma,O,R,\varepsilon} \ \sum_{i} (
\rho_{i}+\sigma_{i}) +Ce^{T}\varepsilon
\nonumber
\\
&&\mbox{s.t.}\qquad \rho_{i}-\sigma_{i} =
y_{i}(R-d_{i})+\varepsilon _{i},\qquad
d_{i} = \|x_{i}-O\|_{2},\qquad i=1,\ldots,n,\hspace*{-25pt}
\nonumber
\\[-8pt]
\\[-8pt]
\nonumber
&&\phantom{\mbox{s.t.}\qquad}(\rho_{i},\sigma_{i},1)\in
S_{3},\qquad i=1,\ldots,n,
\\
&&\phantom{\mbox{s.t.}\qquad}R\ge0,\qquad \varepsilon\ge0.
\nonumber
\end{eqnarray}
This problem is almost a Second Order Cone Program except that the
equality constraints
\begin{eqnarray*}
&& \bigl\{d_{i}=\|x_{i}-O\|_{2}, i=1,\ldots,n\bigr\}
\end{eqnarray*}
are nonlinear (which also makes the problem nonconvex). We use the
first order Taylor expansion iteratively to approximate the nonlinear
equalities by linear ones, which is detailed in the following algorithm
in Section~\ref{sec42}.

\subsection{An iterative algorithm to numerically solve radial
DWD}\label{sec42}

We consider applying the first order Taylor expansion iteratively to
bypass the nonlinearity of the equality constraints and numerically
solve Radial DWD:
\begin{longlist}[]
\item[\textit{Initialization} (\textit{Step} 0):]
Choose an initial center of the separating hypersphere and denote it as
$O^0$ (e.g., the mean or the coordinate-wise median of the $+1$ class
training data), let the initial objective value be $\mathrm{Obj}^{0}=-1$ (an
arbitrary negative number).

\item[\textit{The iteration at Step} $k$:] $k\ge1$.
Apply the first order Taylor expansion on $d_i$ around $O^{k-1}$, that is,
\begin{eqnarray*}
d_{i} &=& \|x_{i}-O\|_{2}
\\
&\approx & \bigl\|x_{i}-O^{k-1}\bigr\|_{2}+\bigl(
\nabla_{O=O^{k-1}} \| x_{i}-O\|_{2}\bigr)^{T}
\bigl(O-O^{k-1}\bigr)
\\
&=& \bigl\|x_{i}-O^{k-1}\bigr\|_{2}-\frac{(x_{i}-O^{k-1})^{T}}{\|
x_{i}-O^{k-1}\|_{2}}
\bigl(O-O^{k-1}\bigr)
:= d^{'}_{i}.
\end{eqnarray*}

Notice that $d^{'}_{i}$ is a linear function of $O$. By substituting
$d_i$ with $d^{'}_i$, the optimization becomes a valid Second Order
Cone Program and could be solved for $O^k$ and $R^k$ using SDPT3.\vspace*{1pt} Let
$\mathrm{Obj}^k$ be the current objective value at step $k$.
\item[\textit{Stop}:] if $|\mathrm{Obj}^{k}-\mathrm{Obj}^{k-1}|<\epsilon$, where $\epsilon$ is a
predetermined precision parameter.
\end{longlist}

At each step $k$, to approximate well the nonlinear terms by using the
first order Taylor expansion, we further confine $O^k$ in a
neighborhood of $O^{k-1}$ (the solution computed from the previous
step) by adding one more constraint: $\|O^k-O^{k-1} \|_2\le\delta
_{k}$, where $ \delta_{k}\in R_{+}$ is called the \emph{step length}
parameter. A small $\delta_{k}$ guarantees the precision of the Taylor
expansion but may slow down the computation. This additional constraint
is a Second Order Cone constraint $(\delta_{k},O^{k}-O^{k-1}) \in
S^{d+1}$ so that we still end up with a valid Second Order Cone Program
at step $k$. In our current data analysis, we choose $\epsilon=
10^{-4}$ and $\delta_{k} = 10^{-3}$. The choice of penalty $C$ will be
revisited after the discussion of Radial DWD optimality conditions in
Section~\ref{sec43}.

\subsection{The dual problem of radial DWD}\label{sec43}
To gain more insights about Radial DWD optimization, it is useful to
give the dual formulation of the Second Order Cone Program (at the
$k$th step).\vspace*{1.5pt} Let $w^{k-1}_{i}=\frac{x_{i}-O^{k-1}}{\|x_{i}-O^{k-1}\|
_{2}}\in R^{d}$, $d^{k-1}_{i}= \|x_{i}-O^{k-1} \|_2$ and they are
functions of $x_{i}$ (since $O^{k-1}$ is computed from the previous
step). After some algebra, we could formulate the dual program at step
$k$ as follows:
%
\begin{eqnarray}
&& \mathrm{Max}_{z} \ \sum_{i}
y_{i}z_{i}d^{k-1}_{i}+
\delta_{k}\biggl(-\biggl\| \sum_{i}
y_{i}z_{i}w^{k-1}_{i}\biggr\|_{2}
\biggr)+2\sum_{i} \sqrt{z_{i}}
\nonumber
\\
&&\mbox{s.t.}\qquad 0\le z_{i} \le C,\qquad i=1,\ldots,n,
\\
&&\phantom{\mbox{s.t.}}\qquad\sum_{i} y_{i}z_{i}
\le0.
\nonumber
\end{eqnarray}
The primal and dual problems can be expressed more compactly in
matrix-vector form. Keep all the defined notation unchanged and denote
$y$ as an $n$-vector of $y_{i}$, $z$ an $n$-vector of $z_{i}$, $\rho$
and $\sigma$ the $n$-vectors of $\rho_{i}$ and $\sigma_{i}$,
respectively, $Y$ an $n$-by-$n$ matrix with $y_{i}$ on the diagonal.
Additionally, let $W_{k-1}=(w^{k-1}_{1},\ldots,w^{k-1}_{n})\in R^{d\times
n}$ with $w^{k-1}_{i}$ defined above, $\triangle^{k-1}_{O}=O-O^{k-1}\in
R^{d}$ and $d_{k-1}=(d^{k-1}_{1},\ldots,d^{k-1}_{n})^{T}\in R^{n}$. Then,
the primal-dual pair becomes
\begin{eqnarray}
&& (\mbox{Primal}) \qquad \mathrm{Min}_{\rho,\sigma,\triangle
^{k-1}_{O},R,\varepsilon} \ e^{T}
\rho+e^{T}\sigma +Ce^{T}\varepsilon
\nonumber
\\
&&\phantom{(\mbox{Primal})}\qquad\mbox{s.t.} \qquad \sigma-\rho+Ry+Y W^{T}_{k-1}
\triangle ^{k-1}_{O}+\varepsilon=Yd_{k-1},
\nonumber
\\[-8pt]
\\[-8pt]
\nonumber
&& \phantom{(\mbox{Primal})}\qquad\phantom{\mbox{s.t.}}\qquad \bigl(
\delta_{k},\triangle^{k-1}_{O}\bigr)\in
S_{d+1},\qquad (\rho _i,\sigma_i,1)\in
S_{3}, \qquad i=1,\ldots,n,
\nonumber
\\
&&\phantom{(\mbox{Primal})}\qquad\phantom{\mbox{s.t.}}\qquad R\ge0, \varepsilon\ge0,
\nonumber
\\
&& (\mbox{Dual}) \qquad \mathrm{Max}_{z} \qquad d^{T}_{k-1}Yz+
\delta _{k}\bigl(-\|W_{k-1}Yz\|_2\bigr)+2e^{T}
\sqrt{z}
\nonumber
\\
&& \phantom{(\mbox{Dual})}\qquad\mbox{s.t.}\qquad 0\le z \le Ce,
\\
&&\phantom{(\mbox{Dual})}\qquad\phantom{\mbox{s.t.}}\qquad y^{T}z \le 0,
\nonumber
\end{eqnarray}
where $\sqrt{z}$ is a $n$-vector with $\sqrt{z_{i}}$ as entries.

One can show the existence of strict feasible solutions to both the
primal and dual problems. Since the primal and the dual are convex, it
follows that the solution of the following optimality conditions are
guaranteed to be optimal or, equivalently, the following equations are
sufficient and necessary optimality conditions:
\begin{eqnarray*}
&& \sigma-\rho+Ry+YW^{T}_{k-1}\triangle^{k-1}_{O}+
\varepsilon=Yd_{k-1},
\\
&& 0<z\le Ce,\qquad \varepsilon\ge0,\qquad (Ce-z)^{T}\varepsilon=0,
\\
&& R\ge0,\qquad y^{T}z\le0,\qquad R\bigl(y^{T}z\bigr)=0.
\\
&&\mbox{Either } W_{k-1}Yz=0\quad \mbox{and}\quad
\bigl\|O-O^{k-1}\bigr\|\le\delta_{k},
\\
&&\mbox{or}\quad \bigl\|O-O^{k-1}\bigr\|_2=\delta _{k}(W_{k-1}Yz)/
\|W_{k-1}Yz\|_2,
\\
&&\rho_{i}=\frac{z_{i}+1}{2\sqrt{z_{i}}} \sigma_{i}=
\frac
{z_{i}-1}{2\sqrt{z_{i}}} \qquad\mbox{for all }i=1,\ldots,n.
\end{eqnarray*}

It is important to note that the optimal radius is strictly positive in
case the 2 training classes are separable and the penalty term $C$ is
large enough, which is shown in Theorem~1 [see the supplementary
materials in \citet{Xiong2015}]. If this is true, we could replace
the condition $\{ R\ge0, y^{T}z\le0, R(y^{T}z)=0\}$ by $\{ R> 0,
y^{T}z = 0\}$. As one will see in Section~\ref{sec44}, this condition gives an
insight to the Radial DWD optimization. Besides, Theorem~1 also implies
that the choice of the penalty parameter $C$ should satisfy the
following: $C(d^{k-1}_{i})^{2}>1$ for all $d_i$, $i\in\{i:y_i=-1\}$.

Solving the primal/dual problem in an ultra high dimension may be
inefficient. To deal with this issue, we first factor the data matrix
$X$ using a \emph{QR decomposition}, for example, $X=QU$ where $Q \in
R^{d\times n}$ has orthonormal columns and $U \in R^{n\times n}$ is an
upper triangular matrix. Then we solve the optimization problem by
replacing $X$ by $U$ and call it a \emph{reduced problem}. The reduced
problem could be solved more efficiently because it shrinks the
intrinsic dimension of the problem from $d$ to the sample size $n$.
Note that it is fairly easy to recover $X$ from $U$ once we solve the
reduced problem. The reduced problem does not change the optimal
solution or the optimal value, which is shown in Theorem~2 [see the
supplementary materials in \citet{Xiong2015}].

\subsection{Interpretation of the radial DWD dual problem}\label{sec44}
Assume that the two classes are separable with a ``proper'' hypersphere
(a hypersphere with nonzero radius $R>0$) so that $y^{T}z=0$ is
obtained at optima. Notice that $y^{T}z=0$ implies
$e^{T}_{+}z_{+}=e^{T}_{-}z_{-}$, where $z_{+}(z_{-})$ is the subvector
of z corresponding to the $+1$ class ($-1$ class) and $e_{+}(e_{-})$ the
corresponding vector of ones. It makes sense to scale $z$ such that
$e^{T}_{+}z_{+}=e^{T}_{-}z_{-}=1$. We can write $z$ as $\eta z^{*}$,
where $\eta$ is a positive scalar and $z^{*}$ satisfies the additional
scaling condition. By maximizing the dual objective function with
respect to $\eta$ for a fixed $z$, we find if $-d_{k-1}^{T}
Yz^{*}+\delta_{k} (\|W_{k-1} Yz^{*} \|_2)>0$,
\begin{eqnarray*}
&& \sqrt{\hat{\eta}}=\frac{e^{T}\sqrt{z^{*}}}{-d^{T}_{k-1}Yz^{*}+\delta
_{k}(\|W_{k-1}Yz^{*}\|_2) }.
\end{eqnarray*}
Equivalently, the dual objective function becomes
%
\begin{eqnarray}\label{46}
&& \mathrm{Max}_{z^{*}}\qquad \frac{(e^{T}\sqrt
{z^{*}})^{2}}{-d^{T}_{k-1}Yz^{*}+\delta_{k}(\|W_{k-1}Yz^{*}\|_2)}.
\end{eqnarray}
Moreover,
\begin{eqnarray*}
&&\!\! d^{T}_{k-1}Yz^{*}- \delta_{k}\bigl(
\|W_{k-1}Yz^{*}\|_2\bigr)
\\
&&\!\!\qquad =\biggl(\sum_{i\in P} d^{k-1}_{i}z^{*}_{i}-
\sum_{i\in N} d^{k-1}_{i}z^{*}_{i}
\biggr) -\delta_{k}\biggl\|\sum_{i\in P}\frac
{x_{i}-O^{k-1}}{d^{k-1}_{i}}z^{*}_{i}-
\sum_{i\in N}\frac
{x_{i}-O^{k-1}}{d^{k-1}_{i}}z^{*}_{i}
\biggr\|_2\!,
\end{eqnarray*}
where $P$ is the index set of the $+1$ class, and $N$ the index set of
the $-1$ class.

Since $\sum_{i\in P} z^{*}_{i}=1$ with $z^{*}_{i}\ge0$, $\sum
_{i\in P} d^{k-1}_{i}z^{*}_{i}$ is a convex combination of
$d^{k-1}_{i},i\in P$, and it can be interpreted as an \emph{average
distance} from the current center of the separating sphere to the $+1$
class data points. A similar interpretation applies for $\sum_{i\in
N} d^{k-1}_{i}z^{*}_{i}$. When the two classes are separable (and
$R>0$), the positive (negative) data points will be located inside
(outside) the separating sphere so that the average distances of
negative points are larger than that of the positive ones, which implies
\begin{eqnarray*}
&& \sum_{i\in P} d^{k-1}_{i}z^{*}_{i}-
\sum_{i\in N} d^{k-1}_{i}z^{*}_{i}<0
\quad \mbox{or}\quad {-}d^{T}_{k-1}Yz^{*}>0.
\end{eqnarray*}
From the above observation, $-d^{T}_{k-1}Yz^{*}+\delta_{k}(\|
W_{k-1}Yz^{*}\|_2)>0$ is true when two classes are separable. Note that
$-d^{T}_{k-1}Yz^{*}$ is a measure of \emph{separability} of the 2
classes and the bigger the absolute value, the bigger the separability.
Meanwhile, $w^{k-1}_{i}=\frac{x_{i}-O^{k-1}}{d^{k-1}_{i}}$ is a vector
of unit Euclidean norm, pointing from the current center to each data
point.\vspace*{1pt} Define the \emph{centroid} of the $+1$ ($-1$) class as the convex
combination of $w^{k-1}_{i}$, $i\in P$ $(\mbox{or}$ $i\in N$,
resp.) under weights $z^{*}_{i}$. Therefore, $\sum_{i\in
N}\frac{x_{i}-O^{k-1}}{d^{k-1}_{i}}z^{*}_{i}-\sum_{i\in P}\frac
{x_{i}-O^{k-1}}{d^{k-1}_{i}}z^{*}_{i}$ is the vector pointing from the
\emph{centroid} of the $+1$ class to the \emph{centroid} of the $-1$ class,
and its Euclidean norm scaled by $\delta_{k}$ is also a measure of
separability. As a consequence, the whole denominator of (\ref{46}) is
positive and is a measure of separability of the 2 classes. To ensure
optima, the dual problem minimizes the separability between classes
divided by the square of the sum of the square roots of the convex weights.

Note that in some situations, the proportions of the 2 classes in the
data set may not reflect the real proportions in a target population
due to sampling bias, or the 2 classes are extremely unbalanced. The
separating boundary tends to be closer to the class with smaller
training sample size. In the case of biased sampling or unbalanced
data, a weighted version of Radial DWD is more appropriate. \citet{QiaoLiu2009} developed a weighting scheme to improve linear DWD and we
follow the same line to set up weighted Radial DWD, by optimizing the
following objective function:
%
\begin{eqnarray}
&&\sum_{i} w(y_{i})\biggl\{
\frac{1}{r_i}+Ce^{T}\varepsilon\biggr\}
\end{eqnarray}
subject to the same set of constraints defined before. Note that
$w(y_{i})$ is the \emph{weight} associated with the $i$th training
data point and it only depends on the class label~$y_i$. In our\vspace*{1pt} data
analysis we use
$w(+1)=\frac{n_{-}}{n_{+}+n_{-}}; w(-1)=\frac{n_{+}}{n_{+}+n_{-}}$ as
default. The above discussion about the Radial DWD optimization could
be easily generalized to the case when weights are applied.

\section{Conclusion}\label{sec5}
In this article we have proposed a nonlinear binary classifier, Radial
DWD, for the virus hunting data analysis, where the virus positive
class is surrounded by the negative class in very diverse radial
directions. Because of the nonlinearity of classes, linear methods,
including MD, LASSO logistic regression, SVM and DWD, perform poorly
with high classification error. Meanwhile, kernel SVM shows a very
limited improvement over its linear counterpart in high dimensions.
Since standard nonlinear methods, including kernel methods, require a
type of ``data richness,'' that is not present in the virus hunting
problem. In particular, they work well in situations (such as all the
usual machine learning examples) where training data can be found in
all of the various regions where the test data will appear. But in our
particular data context, that completely breaks down, so all the
classical nonlinear methods fare just as poorly as the linear ones. By
using a much more appropriate spherical separating boundary, Radial DWD
shows both low false positive and low false negative classification
error. These are shown by real data analysis and simulation studies.
Its computation through solving a sequence of Second Order Cone
Programs is efficient, even with high-dimensional data.

We believe Radial DWD will be applicable in some settings beyond virus
hunting. This will happen in classification contexts where there is one
class with relatively small variation, and the other with much larger
variation tending toward a number of quite divergent directions. For
example, cancer is a disease of sometimes massive disruption of the
genome, and these disruptions can go in many diverse directions, while
the normal genome is far more stable. Another potential for this
methodology comes in imaging bones and cartilage, where the normal
population is relatively homogeneous, but severe wear and other types
of abnormalities can go in many directions in the image space.

\section*{Acknowledgments}
The authors would like to thank the Editor, Associate Editor and
referees for their insightful, constructive comments.

\begin{supplement}[id=suppA]
\stitle{Supplement to: ``Virus hunting'' using Radial Distance Weighted Discrimination.}
\slink[doi]{10.1214/15-AOAS869SUPP} 
\sdatatype{.pdf}
\sfilename{aoas869\_supp.pdf}
\sdescription{In the supplementary materials, we first introduce some
useful biology background for virus detection in Section~1, DNA
alignment process in Section~2, and then discuss the insights of the
Dirichlet distribution in Section~3. Real data examples and simulation
studies are included in Sections~4 and 5, respectively. Theorems and
proofs are in Section~6.}
\end{supplement}


\printaddresses
\end{document}